\documentclass[a4paper,11pt]{article}
\usepackage[dvips]{graphicx}
\usepackage[english]{babel}

\usepackage{amsmath}
\usepackage{latexsym}
\usepackage{amsmath}
\usepackage[all]{xy}

\begin{document}

\title{Comments on scalar-tensor
representation of  non-locally corrected gravity.}

\author{N. A. Koshelev \\[5mm]
Ulyanovsk State University,\\
Leo Tolstoy str 42,  432970, Russia \\ \\
\it e-mail: koshna71@inbox.ru \\
}

\date{}
\maketitle \abstract{The scalar-tensor representation of
nonlocally corrected gravity is considered. Some special solutions
of the vacuum background equations were obtained that indicate to
the nonequivalence of the initial theory and its scalar-tensor
representation. }

\section{Introduction}

Recent  observations have shown that the universe is presently
accelerating. Most attempts to explain this acceleration involve
the introduction of dark energy, as a source of the Einstein field
equations. Standard $ \Lambda CDM $ model, in which dark  energy
is considered as a small positive cosmological constant, provides
a very good fit to the supernovae data \cite{Astier} as well as
CMB measurements \cite{WMAP5} and observations of large scale
structure \cite{SDSS} , but the small and fine-tuned value of the
cosmological constant cannot be explained within current particle
physics \cite{Weinberg}. As a result a many other  cosmological
models have been proposed to giving a dynamical origin to dark
energy. Pure phenomenologically one can consider  dark energy as a
perfect fluid with a sufficiently negative pressure
\cite{Choudhury_Padmanabhan}. There is also a large class of
scalar field models in the literature including quintessence
\cite{SWZ} and phantom \cite{Caldwell} fields.

Scalar-tensor theories of gravity called an attention in
connection with attempts to give an geometrical explanation to a
dark energy phenomenon \cite{Scalar-tensor_0},
\cite{Scalar-tensor_1}, \cite{Scalar-tensor_f},
\cite{Scalar-tensor_2}. They are conformally coupled with
intensively studied  $f(R)$ generalized gravity theories in metric
and Palatini formalism. Although several $f(R)$ modified gravity
models have been proposed which realize the correct cosmological
evolution and satisfy solar system tests, for the current moment
there are some considerable problems in construction of available
models of dark energy on the basis of theories of this type
\cite{Sotiriou_Faraoni} (see also ref. \cite{Kobayashi_Maeda}).

Several papers have appeared recently containing scalar-tensor
representation of nonlocal theories \cite{Nojiri_Odintsov_0708},
\cite{JNOSTZ}, \cite{Koivisto_0803}, \cite{Odintsov_0908}. The
simple example of the nonlocal action \cite{Deser_Woodard} is
\begin{equation}
\label{action1} S=\int d^4
x\sqrt{-g}\left\{\frac{R}{2\kappa^2}(1+f( \Box ^{-1} R))
+L_m\right\} .
\end{equation}
where $R$ is the Ricci scalar, $\Box $  is the d'Alembertian, and
$L_m$ is the Lagrangian of matter. More general action involving
$m$ different powers $\Box ^{-1}$ acting on $R$ also can be
considered \cite{JNOSTZ}.

The standard approach is based on introducing the Lagrange
multiplier $\xi$ and rewriting the initial theory in the local
form
\begin{equation}
\label{action2} S=\int d^4
x\sqrt{-g}\left\{\frac{1}{2\kappa^2}\left\{R(1+f(\phi)) +\xi (\Box
\phi -R)\right\} + L_m \right\} .
\end{equation}

Validity of such approach raises the doubts. On the one hand, the
constraint equation
\begin{equation}
\label{constraint} \Box \phi -R =0
\end{equation}
has the formal solution $\phi =\Box ^{-1}R$. Substituting the
above solution into (\ref{action2}) one reobtains (\ref{action1}).

On the other hand, unlike classical mechanics, constraint
(\ref{constraint}) contains the second order time derivatives.
Besides, this constraint equation does not allow to determine $
\varphi (R)$ uniquely, since it remain valid after the
transformation $ \varphi \rightarrow \varphi + \varphi_D $, where
function $ \varphi_D (x)$ is an arbitrary solution of the
d'Alembert's equation. For the proof of equivalence of theories it
is necessary to check up equivalence of the dynamical equations
that has not been done earlier. The purpose of this work is a more
detailed analysis of the relationship between models
(\ref{action1}) and (\ref{action2}).

\section{The basic equations}

In this section we shortly describe some equivalent form of action
(\ref{action2}) and the basic equations. Integrating by parts and
neglecting the boundary terms, the action (\ref{action2}) can  be
rewritten as

\begin{equation}
S=\int d^4 x\sqrt{-g}\left\{\frac{1}{2\kappa^2}\left\{R(1+f(\phi)
-\xi) - \xi_{,\alpha} \phi^{,\alpha}\right\} +L_m \right\}.
\end{equation}
It is convenient to introduce a new scalar field $ \Psi = f (\phi)
- \xi $, which is one nonminimally coupled to gravity in physical
 (Jordan) frame. The action then becomes \cite{Koivisto_0803}

\begin{equation}
\label{action3} S=\frac{1}{2\kappa^2}\!\int \! d^4
x\sqrt{-g}\left\{(1\!+\Psi)R -\! f'(\phi) \phi_{,\alpha}
\phi^{,\alpha}\!+ \Psi_{,\alpha} \phi^{,\alpha}\!
+2\kappa^2L_m\right\}\!.
\end{equation}
The Einstein frame action can be obtained by conformal
transformation such that

\begin{equation}
\label{Weyl} \tilde{g}_{\mu\nu}\equiv (1+\Psi)g_{\mu\nu}\equiv
e^{\lambda}g_{\mu\nu}.
\end{equation}
Then, one can obtain \cite{Koivisto_0803}:

\begin{equation}
S=\frac{1}{2\kappa^2}\int d^4 x\sqrt{-\tilde{g}}\left\{\tilde{R}
-\frac{3}{2} \lambda_{\widetilde{,\alpha}} \lambda
^{\widetilde{,\alpha}} -
e^{-\lambda}f'(\phi)\phi_{\widetilde{,\alpha}} \phi
^{\widetilde{,\alpha}} \right. \nonumber
\end{equation}
\begin{equation}
\left.+\lambda_{\widetilde{,\alpha}} \phi ^{\widetilde{,\alpha}}
+2\kappa^2e^{-2\lambda}L_m(\tilde{g}e^{-\lambda}) \right\}.
\end{equation}
Applying the field  redefinition

\begin{equation}
\label{redefinition} \lambda =\frac{1}{3}\phi - \sqrt{
\frac{2}{3}}\varphi ,
\end{equation}
yields:

\begin{equation}
\label{action4} S=\frac{1}{2\kappa^2}\int d^4 x\sqrt{-\tilde{g}}
\left\{\tilde{R} - \varphi_{\widetilde{,\alpha}} \varphi
^{\widetilde{,\alpha}} -\left( e^{\sqrt{ \frac{2}{3}}\varphi
-\frac{1}{3}\phi} f'(\phi) -\frac{1}{6} \right) \phi
_{\widetilde{,\alpha}} \phi^{\widetilde{,\alpha}} \right.
\nonumber
\end{equation}
\begin{equation}
\left.+2\kappa^2 e^{ \frac{2\sqrt{2}}{\sqrt{3}}\varphi
-\frac{2}{3}\phi} L_m (\tilde{g}e^{\sqrt{ \frac{2}{3}}\varphi
-\frac{1}{3}\phi}) \right\}.
\end{equation}
Let's notice that the above definition (\ref{redefinition}) is
chosen slightly different from used in \cite{Koivisto_0807} to
provide more evident transition to the limiting case $f (\phi) =
f_0 = const$.

It should be emphasized that in the limiting case $f (\phi) = f_0$
this model reduce to general relativity with one canonic, one
phantom scalar field and the nonminimally coupled to gravity
matter. The correct limiting transition will occur only if the
additional condition $ e^{\frac{1}{3}\phi - \sqrt{
\frac{2}{3}}\varphi} \equiv 1+f_0$ is added. In this case
contributions of scalar fields compensate each other, and the
matter coupled to gravity minimally.

Variation of action (\ref{action3}) with respect to the metric
$g_{\mu\nu}$ gives the modified Einstein equations

$$
\frac{1}{2}R(1+\Psi) g_{\mu\nu} - R_{\mu\nu}(1+\Psi)- \frac{1}{2}
f'\phi_{,\alpha} \phi^{,\alpha} g_{\mu\nu} + \frac{1}{2} \Psi
_{,\alpha} \phi^{,\alpha} g_{\mu\nu} + f' \phi_{,\mu} \phi_{,\nu}
$$
\begin{equation}
\label{gr_eq} -\frac{1}{2}(\Psi_{,\mu}\phi_{,\nu} + \Psi_{,\nu}
\phi_{,\mu})-g_{\mu\nu}\Psi_{;\alpha}^{~\alpha} + \Psi_{;\mu\nu}
+\kappa^2 T_{\mu\nu} =0,
\end{equation}
where $T_{\mu\nu}$ is the energy-momentum tensor of the fluid
matter.

By the variation over $\phi$ and $\Psi$, one can  obtain the
scalar field equations \cite{Nojiri_Odintsov_0708,Koivisto_0803}

\begin{equation}
\Psi_{;\alpha}^{~\alpha} = f''(\phi)(\partial \phi)^2 +
2f'(\phi)R,
\end{equation}
\begin{equation}
\phi_{;\alpha}^{~\alpha} =R.
\end{equation}

Consider now the spatially flat Friedmann-Robertson-Walker
Universe, described by the action (\ref{action3}) with the line
element

\begin{equation}
ds^2 = -dt^2 +a^2(t)dx_idx^i.
\end{equation}
The Friedmann equations can be written as  \cite{JNOSTZ}:

\begin{equation}
3H^2(1+\Psi) =\kappa^2\rho +\frac{1}{2}\left(f' (\phi)
\dot{\phi}^2 -\dot{\Psi}\dot{\phi}\right) -3H\dot{\Psi},
\end{equation}
\begin{equation}
- (2\dot{H} +3H^2)(1+\Psi) =\kappa^2p +\frac{1}{2} \left(f'(\phi)
\dot{\phi}^2 -\dot{\Psi}\dot{\phi} \right)+\ddot{\Psi}
+2H\dot{\Psi},
\end{equation}
where $H=\frac{\dot{a}}{a}$ is the Hubble rate, and energy density
$ \rho$ and pressure $p$  of the background perfect fluid are
satisfy the continuity equation

\begin{equation}
\dot{\rho} +3H(1+w)\rho =0.
\end{equation}

The background field equations are
\begin{equation}
\ddot{\phi} +3H\dot{\phi} = -6(\dot{H} +2H^2),
\end{equation}
\begin{equation}
\ddot{\Psi} +3H\dot{\Psi} =f''(\phi)\dot{\phi}^2
-12f'(\phi)(\dot{H} +2H^2).
\end{equation}

\section{The analysis of the background equations at $f '(\varphi)\equiv $0}

In what follows we consider the background equations in the simple
special case $f '(\varphi)\equiv 0$. The equations of motion for
the homogeneous background scalar fields and scale factor $a(t)$
are simplify to

\begin{equation}
3H^2(1+\Psi) =\kappa^2\rho -\frac{1}{2}\dot{\Psi}\dot{\phi}
-3H\dot{\Psi},
\end{equation}
\begin{equation}
- (2\dot{H} +3H^2)(1+\Psi) =\kappa^2p
-\frac{1}{2}\dot{\Psi}\dot{\phi} +\ddot{\Psi} +2H\dot{\Psi},
\end{equation}
\begin{equation}
\ddot{\phi} +3H\dot{\phi} = -6(\dot{H} +2H^2),
\end{equation}
\begin{equation}
\ddot{\Psi} +3H\dot{\Psi} =0,
\end{equation}
where $p= p (\rho)$.

In absence of matter we obtain a set of the equations describing
dynamics of three variables $ \phi $, $ \Psi $ and $H $.

\begin{equation}
\label{absence_begin} 3H^2(1+\Psi)
=-\frac{1}{2}\dot{\Psi}\dot{\phi} -3H\dot{\Psi},
\end{equation}
\begin{equation}
- 2\dot{H} (1+\Psi) = -\dot{\Psi}\dot{\phi} -4H\dot{\Psi},
\end{equation}
\begin{equation}
\ddot{\phi} +3H\dot{\phi} = -6(\dot{H} +2H^2),
\end{equation}
\begin{equation}
\label{absence_end} \ddot{\Psi} +3H\dot{\Psi} =0.
\end{equation}
This system has the trivial solution

\begin{equation}
\label{solution1} \phi(t)=\phi_0, ~~~\Psi(t) =\Psi_0,~~~H(t)=0,
\end{equation}
that corresponds to the Minkowski metric. However other solutions
are also available.

Assuming $ \dot {H} \equiv 0$, the system of equations
(\ref{absence_begin}) - (\ref{absence_end}) is reduced to

\begin{equation} 3H^2(1+\Psi) =-\frac{1}{2}\dot{\Psi}\dot{\phi}
-3H\dot{\Psi},
\end{equation}
\begin{equation}
0 = \dot{\Psi}\dot{\phi} + 4H\dot{\Psi},
\end{equation}
\begin{equation}
\ddot{\phi} +3H\dot{\phi} = -12H^2,
\end{equation}
\begin{equation}
\ddot{\Psi} +3H\dot{\Psi} =0,
\end{equation}
with the de Sitter solution

\begin{equation}
\label{solution2} \Psi =Ae^{-3H_0t} - 1, ~~~~\phi =-4H_0t +
\phi_0, ~~~~ H=H_0,
\end{equation}
where $A$, $\phi_0$, $H_0$ are arbitrary constants.

Analogously, considering the case $ \dot {\phi} \equiv 0$, one can
obtain

\begin{equation}
\label{solution3} \Psi =\frac{A}{(C+2t)^{1/2}} -1,~~~\phi
=\phi_0,~~~ H = \frac{1}{C+2t}.
\end{equation}

It is remarkable, that the initial theory (\ref{action1}) in the
case under consideration reduce to

\begin{equation}
S=\frac{1}{2\kappa^2}\int d^4 x\sqrt{-g}\left\{
\frac{R}{2\kappa^2} (1+f_0) +L_m\right\} .
\end{equation}

This is the action of general relativity with  a renormalized
gravitational constant. The only homogeneous and isotropic
solution of the corresponding vacuum Einstein field equations is
the Minkowski metric. Hence, background solutions
(\ref{solution2}) and (\ref{solution3}) are artifacts of the
scalar-tensor representation.

The existence of these solutions show that generalized Einstein
gravity theories (\ref{action1}) and (\ref{action2}) are not
equivalent even at $f '(\phi)\equiv 0$. In the scalar-tensor
representation there are new solutions for the metrics which are
unavailable in the initial theory.

\section{Conclusion}

We consider the general equations of motion and the background
equations for scalar-tensor theory (\ref{action2}). We have
obtained particular solutions of the background equations, which
show the nonequivalence  of theories (\ref{action1}) and
(\ref{action2}) even in the limiting case $f ' (\phi) \equiv 0$.
An available problem is the existence of redundant solutions in
the scalar-tensor representation of nonlocally corrected gravity.
It remains an open question whether constraints can also be
satisfied to disregard unphysical solutions.

\end{document}